%% file: ATLAS.tex
\documentclass{aa}  

\usepackage{graphicx}
\usepackage{txfonts}

\usepackage{natbib}
\usepackage{dcolumn,url,longtable}
\usepackage{amssymb}
\usepackage{color}
\usepackage{tikz}
\usetikzlibrary{shapes,arrows}
\usepackage{pdflscape}

\newcolumntype{d}{D{.}{.}{-1}}

\begin{document} 

  \title{Asteroid models reconstructed from ATLAS photometry}

  \author{J.~\v{D}urech         \inst{1}        \and
          J.~Tonry              \inst{2}        \and
          N.~Erasmus            \inst{3}        \and    
          L.~Denneau            \inst{2}        \and
          A.~N.~Heinze          \inst{2}        \and
          H.~Flewelling         \inst{2}        \and    
          R.~Van\v{c}o          \inst{1}
          }

  \institute{Astronomical Institute, Faculty of Mathematics and Physics, Charles University, V~Hole\v{s}ovi\v{c}k\'ach~2, 180\,00 Prague~8, Czech Republic\\
             \email{durech@sirrah.troja.mff.cuni.cz}
              \and
              Institute for Astronomy, University of Hawaii, Honolulu, HI, USA
              \and
              South African Astronomical Observatory, Cape Town, 7925, South Africa
             }

  \date{Received ?; accepted ?}

  \abstract
  {The Asteroid Terrestrial-impact Last Alert System (ATLAS) is an all-sky survey primarily aimed at detecting potentially hazardous near-Earth asteroids. Apart from the astrometry of asteroids, it also produces their photometric measurements that contain information about asteroid rotation and their shape.}
  {To increase the current number of asteroids with a known shape and spin state, we reconstructed asteroid models from ATLAS photometry that was available for approximately 180,000 asteroids observed between 2015 and 2018.}
  {We made use of the light-curve inversion method implemented in the Asteroid@home project to process ATLAS photometry for roughly 100,000 asteroids with more than a hundred individual brightness measurements. By scanning the period and pole parameter space, we selected those best-fit models that were, according to our setup, a unique solution for the inverse problem.}
  {We derived $\sim$\,2750 unique models, 950 of them were already reconstructed from other data and published. The remaining 1800 models are new. About half of them are only partial models, with an unconstrained pole ecliptic longitude. Together with the shape and spin, we also determined for each modeled asteroid its color index from the cyan and orange filter used by the ATLAS survey. We also show the correlations between the color index, albedo, and slope of the phase-angle function.}
  {The current analysis is the first inversion of ATLAS asteroid photometry, and it is the first step in exploiting the huge scientific potential that ATLAS photometry has. ATLAS continues to observe, and in the future, this data, together with other independent photometric measurements, can be inverted to produce more refined asteroid models.}


  \maketitle

  \section{Introduction}

    Asteroids are important, not only as a subject of scientific studies, but also because some of them are potentially hazardous and can collide with Earth. There are several sky surveys that map the population of asteroids and near-Earth asteroids (NEAs), in particular. One of them is the Asteroid Terrestrial-impact Last Alert System (ATLAS), whose main purpose is to detect dangerous NEAs. The system currently consists of two Schmidt telescopes, which are located at Haleakal\={a} and Mauna Loa in Hawaii. The telescopes are identical; they both have a 65-cm primary mirror, $f/2.0$, and the field of view is $7.5^\circ$. They cover the whole visible sky for one night down to $\sim 19$\,mag with 30\,s exposures \citep{Ton.ea:18, Ton.ea:18b, Smi.ea:20, Hei.ea:18}. 

    Apart from astrometric measurements that are used for asteroid orbit determination, ATLAS also measures asteroid brightness. As such, for every detected asteroid, we have a sparsely sampled light curve. Our study aims to use this ATLAS asteroid photometry to reconstruct the basic shapes of asteroids and their rotation properties by the light curve inversion method of \cite{Kaa.ea:01}.

  \section{ATLAS photometric data}

    We processed all available photometry for about 180,000 asteroids observed from June 2015 to October 2018. Most of the observations were done in orange ($o$, 560--820\,nm) and cyan ($c$, 420--650\,nm) filters. For each observation, we computed the heliocentric and geocentric coordinates of the asteroid, which were needed to compute the viewing and illumination geometry, using the Miriade\footnote{\url{http://vo.imcce.fr/webservices/miriade/}} system of IMCCE \citep{Ber.ea:09}. The apparent brightness was normalized to a distance of 1\,AU from the asteroid to the Sun and to the Earth.

    There were some outliers in the data, which are clearly visible on phase curves, when we plotted the brightness as a function of the solar phase angle. We had to remove these outliers before using the data for inversion. To do so, we used the same approach as \cite{Han.ea:11} and fit a linear-exponential phase function to data and computed the root mean square (rms) residual. Then we removed all measurements that differed from the fitted phase function by more than a given limit. After some tests and the visual inspection of several asteroids, we set up this limit to 2.5 rms, which seemed to be a good compromise that removed most of the outliers and did not drastically reduce the number of data points. On average, this removed about 15\% of the data points.
    
    After cleaning the photometry, we selected only those asteroids with at least 100 observations in order to have enough data for the inversion. This way, we obtained a sample of about 98,000 asteroids to which we applied the light curve inversion method of \cite{Kaa.ea:01}.

  \section{Shape and spin reconstruction with light curve inversion}

    For the inversion of ATLAS photometry, we used the same approach as employed with Lowell Observatory photometry \citep{Dur.ea:16}, Lowell data combined with WISE data \citep{Dur.ea:18c}, Lowell and Gaia DR2 data \citep{Dur.ea:19}, and Gaia data alone \citep{Dur.Han:18}. The core of the problem was to find the correct sidereal rotation period. To do so, we started the inversion algorithm from many different initial trial periods and it converged to different local minima from which the global minimum was selected. We used full convex shape inversion (running at Asteroids@home project\footnote{\url{http://asteroidsathome.net}}) and also a simpler ellipsoidal shape model that ran at the computational cluster Tiger at the Astronomical Institute of Charles University. The detailed description of the processing pipeline and reliability tests can be found in \cite{Han.ea:11} and \cite{Dur.ea:18c}.

    The most time-consuming part of the modeling is the search for the sidereal rotation period. Although ATLAS data are, for many asteroids, dense enough to apply Fourier-based algorithms in order to find periodicity in the signal \citep{Era.ea:20}, our aim was different. Instead of determining the synodic rotation period, we wanted to derive the full spin state, including the sidereal rotation period with the relative precision on the order of $10^{-5}$, which corresponds to knowing the exact number of rotations in the time interval covered by the observations. So we scanned a range of periods from 2--1000\,h with our full physical model. From the whole sample of almost a hundred thousand asteroids, a reliable sidereal period was determined for only about five thousand. So, for the majority of asteroids, their photometric data were not sufficient in number nor in accuracy to uniquely determine the sidereal rotation period; apart from the correct solution, there were other possible periods with an acceptable fit. Even if such periods were almost identical from the point of view of period determination -- a relative difference on the order of $10^{-3}$ -- the corresponding pole directions and shapes were very different, so we could not consider these as unique models.

    To obtain more models, we used the Lightcurve Database (LCDB) of \cite{War.ea:09}, the version from November 20, 2018, where rotation periods derived from independent light curves are stored. We used these periods as a priori information, which enabled us to shrink the period parameter space significantly. Because the periodograms were computed on the whole 2--1000\,h interval for all asteroids, we just constrained the solution to some interval around the known period $P_\text{LCDB}$. The reliability of periods in the LCDB is marked with an uncertainty code $U$, with higher values meaning a more reliable period. We used only U values of 2 and higher and limited our search intervals to $\pm 5\%$ around $P_\text{LCDB}$ for $U = 3$ or $3-$, to $\pm 10\%$ for $U = 2+$, and to $\pm 20\%$ for $U = 2$. 
  
  \section{Results}

    In total, we obtained about 2750 unique asteroid models that passed our reliability tests. Out of these, about 950 asteroids already had a model that was reconstructed from independent data and was published. The remaining 1800 cases are new asteroid models,  the parameters of which we publish for the first time. They are listed in Table~\ref{tab:models} (705 full models), Table~\ref{tab:models_interval} (191 models derived with a priori information about their rotation period), Table~\ref{tab:models_partial} (671 partial models with unconstrained pole ecliptic longitude), and Table~\ref{tab:models_interval_partial} (225 partial models with a priori period information). The full models from Tables~\ref{tab:models} and \ref{tab:models_interval} are also available on the Database of Asteroid Models from Inversion Techniques \citep[DAMIT,][]{Dur.ea:10}.\footnote{\url{https://astro.troja.mff.cuni.cz/projects/damit}} In most cases, there were two possible pole solutions with about the same pole latitude $\beta$ and pole longitudes $\lambda$ about $180^\circ$ apart. This is a consequence of the problem's symmetry because observations limited to the ecliptic plane always lead to this $\lambda \pm 180^\circ$ ambiguity \citep{Kaa.Lam:06}. 
    
    For so-called partial models in Tables~\ref{tab:models_partial} and \ref{tab:models_interval_partial}, ATLAS data were not sufficient to uniquely reconstruct their spin axis direction. However, their pole latitude was limited to an interval of $\beta \pm \Delta\beta$, which we report. Additional data are needed to reconstruct full models for these objects uniquely.

    \subsection{Error estimation}

      To estimate errors of the derived model parameters, we assumed that individual ATLAS measurements were independent, and we used a bootstrap method. We created one hundred bootstrapped data samples for every full and partial model. The original photometric data were randomly sampled with a replacement; the number of data points observed in each filter remained the same. We ran the convex light curve inversion for each bootstrap sample again  using the spin parameters $(\lambda, \beta, P)$ of the original model as initial values. The inversion algorithm converged to a different set of spin, phase function, and color parameters. The uncertainty of the pole ecliptic longitude $\sigma_\lambda$ and latitude $\sigma_\beta$, the sidereal rotation period $\sigma_P$, and the color index $\sigma_{c-o}$ was determined as a standard deviation of the sample for one hundred values. These uncertainties are also listed in Tables~\ref{tab:models} to \ref{tab:models_interval_partial}.      

    \subsection{Period comparison}
\begin{figure}[t]
       \includegraphics[width=\columnwidth]{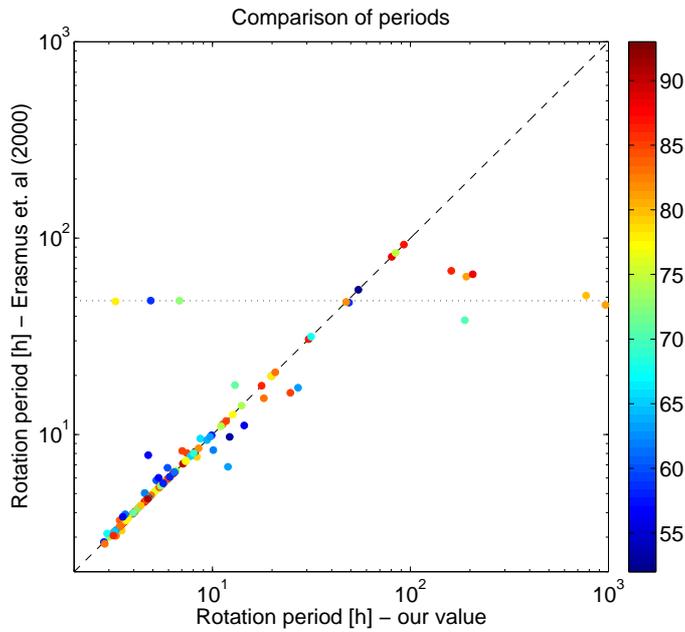}
       \caption{Comparison of rotation periods of 134 asteroids that are in our sample and also in that of \cite{Era.ea:20}. Colors represent the confidence of period determination (in percent) according to \cite{Era.ea:20}. The dotted horizontal line marks a rotation period of 48 hours.}  
       \label{fig:periods_comparison}
      \end{figure}

      \cite{Era.ea:20} used the same ATLAS data set to determine the colors and rotation periods of members of selected asteroid families. There were only 134 asteroids for which we could compare our values with those in \cite{Era.ea:20}. We only used the periods determined uniquely on the whole 2--1000\,h interval, that is to say the values from Tables~\ref{tab:models} and \ref{tab:models_partial}. The comparison of the periods is shown in Fig.~\ref{fig:periods_comparison}. The color coding represents the confidence of the period determination according to \cite{Era.ea:20}. There were 95 asteroids for which the periods were the same within the $3\sigma$ interval. Of the remaining 39 asteroids, all of the periods reported by \cite{Era.ea:20} were either $\pm 2$, $\pm 1$, $\pm 0.5$, or $\pm 0.25$-day aliases of the period we report here or the periods extracted were close to 48 hours. Aliasing ambiguity in the extracted (by Lomb-Scargle periodogram) rotation period and incorrectly resolving periods close to 12, 24, or 48 hours is a common problem with ATLAS data due to the one-day cadence of the ATLAS observations. 
   
    \subsection{Phase functions}

      \begin{figure}[t]
        \includegraphics[width=\columnwidth]{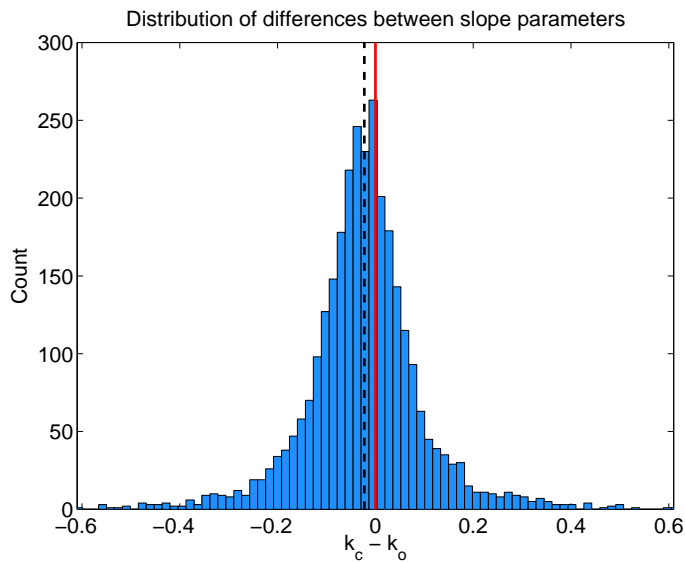}
        \caption{Histogram of differences between slope parameters $k_c$ and $k_o$ for cyan and orange filters. The red vertical line emphasizes that the distribution is not centered at zero, but shifted to the left. The black dashed line marks the mean value of $k_c - k_o = -0.023$.}
        \label{fig:slope_difference_hist}
      \end{figure}

      Similarly as in our previous works, we described the light-scattering properties of our convex models as a combination of Lomell-Seeliger and Lambert scattering \citep{Kaa.ea:01}. The bi-directional reflectance is 
      \begin{equation}
        S(\mu, \mu_0, \alpha) = f(\alpha) \, \mu \mu_0 \, \left( \frac{1}{\mu + \mu_0} + c \right)\,,
      \end{equation}
      where $\mu$ and $\mu_0$ are cosines of emission and incidence angles, respectively, $c = 0.1$ is a constant of the Lambert part, and $f(\alpha)$ is the phase angle function 
      \begin{equation}
        \label{eq:phase_function}
        f(\alpha) = A_0 \exp \left( -\frac{\alpha}{\cal{D}} \right) + k \alpha + 1\,,
      \end{equation}
      where $A_0$, $\cal{D}$, and $k$ are parameters that are to be optimized and $\alpha$ is the solar phase angle. This scattering model is mathematically simple, the phase angle function $f(\alpha)$ does not depend on $\mu$ or $\mu_0$, and the model is flexible enough to fit our data. 

      We assumed that the phase function is the same for both orange and cyan colors, that is to say that the scattering parameters in eq.~(\ref{eq:phase_function}) do not depend on the wavelength. This is not exactly true because the phase function is, in fact, dependent on the wavelength; the effect is known as ``phase reddening'' \citep{Mil.ea:76, Lum.Bow:81b}. To test if ATLAS data show some information about phase reddening, we took all unique models and repeated the inversion, starting from the final spin parameters, with a modified phase function. We assumed the parameters $A_0$ and $\cal{D}$ are again the same for both colors, but that the linear parameter $k$ can be different. So instead of one value of $k$, we had two independent parameters $k_c$ and $k_o$ for cyan and orange filters. Figure~\ref{fig:slope_difference_hist} shows a histogram of differences $k_c - k_o$ between slope parameters for all models. The distribution is not symmetric around zero; there are more models with negative $k_c - k_o$, meaning that statistically $k_\mathrm{c} < k_\mathrm{o}$. Because values of the slope parameter cannot be positive (this is constrained during optimization), then $|k_c| > |k_o|$. This means that the orange slope is not as steep as the cyan one on average; so for larger phase angles, the brightness is higher in orange than in cyan, which agrees with the concept of phase reddening. However, this behavior can only be shown for the whole sample statistically; individual cases can behave differently and numerical values of $k_c$ and $k_o$ are sensitive to outliers in the input photometry.

      The mean value of $k_c - k_o$ from our sample is $-0.023$ (Fig.~\ref{fig:slope_difference_hist}). If we take the effective wavelengths of $c$ and $o$ filters as 533\,nm and 679\,nm, respectively \citep{Ton.ea:18}, it gives the mean phase reddening coefficient of $0.00027\,\text{mag} / \text{deg} / 100\,\text{nm}$, which is about four times smaller than typical values of $0.001\,\text{mag} / \text{deg}$ for the $B - V$ color (which is also an interval of about 100\,nm) reported by \cite{Lum.Bow:81b}.
      
\begin{figure}[t]
        \includegraphics[width=\columnwidth]{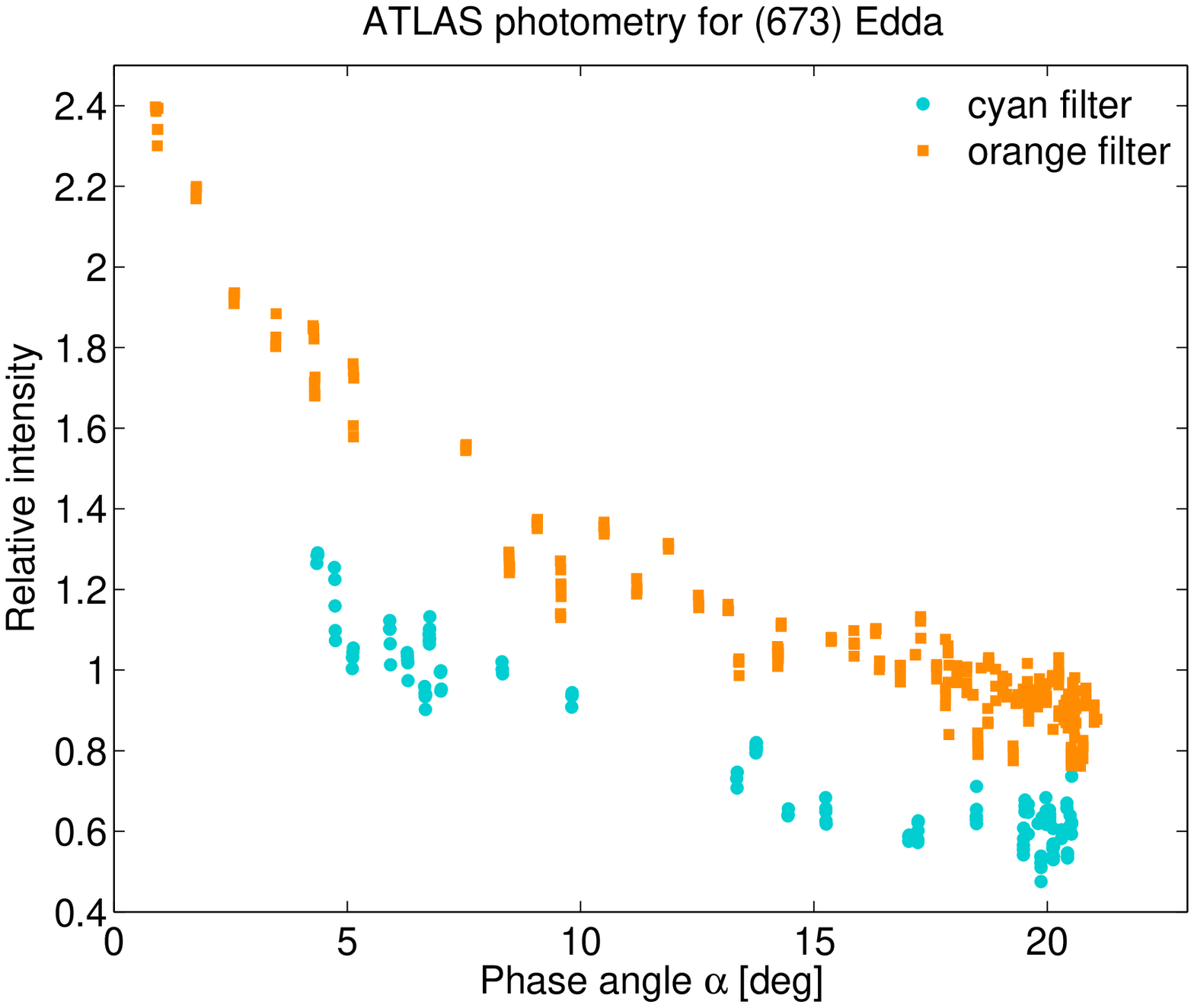}

        \includegraphics[width=\columnwidth]{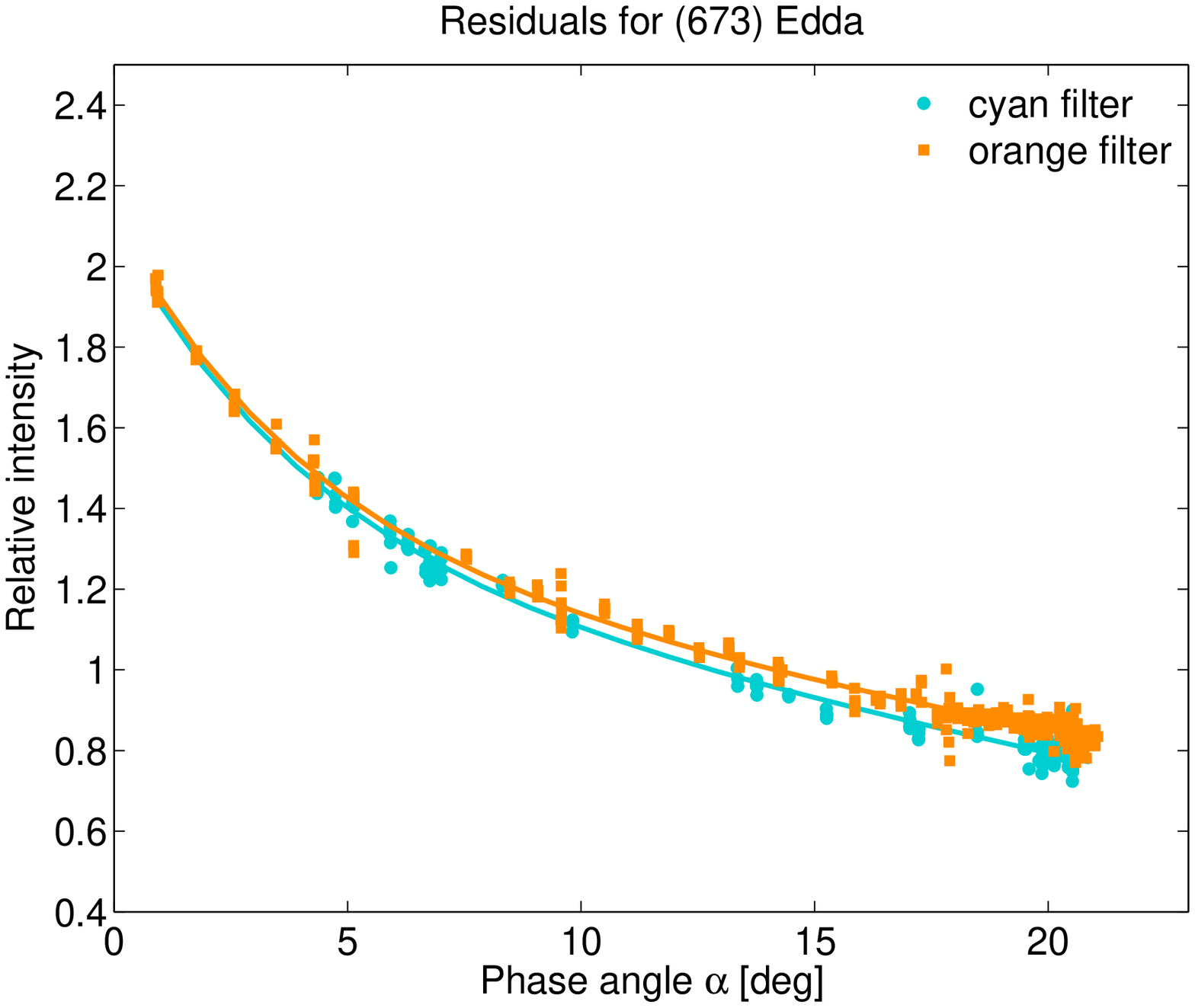}
        \caption{Phase function of asteroid (673)~Edda. The top panel shows the original measurements in cyan and orange filters; the brightness (relative flux) was normalized to a unit distance from the Sun and Earth, and outliers were removed. The vertical shift between two filters corresponds to the color index $c - o = 0.402$\,mag. On the bottom panel, the rotation and geometry effects were removed and the solid curves correspond to phase-angle functions according to eq.~(\ref{eq:phase_function}) with different slope parameters $k_c$ and $k_o$. The scatter of points around the mean phase curves corresponds to residuals of the fit.}
        \label{fig:phase_function}
      \end{figure}

      An example of an asteroid for which the modeling with two slope parameters gives significantly better results than with one slope parameter is (673)~Edda, which is shown in Fig.~\ref{fig:phase_function}. The model from ATLAS data agrees with that of \cite{Mar.ea:19}. We show original ATLAS data in two colors. The color index was determined to $c - o = 0.402$\,mag and was modeled as a free scale factor between the data sets in two filters, or equivalently as an offset in magnitude scale. The spread of points around the mean phase curve is caused by shape effects and photometric errors. After removing the shape and geometry effects, the difference in phase curves for orange and cyan bandpasses becomes visible. The residuals dropped from 0.0286 and 0.0274\,mag for $c$ and $o$ data with a single slope parameter to 0.0274 and 0.0262\,mag when two independent slope parameters were used. 

      In general, splitting photometry into two colors and fitting them with different slopes improves the fit, but usually not very significantly. In most cases, the fit with one slope parameter $k$ is about the same as with two independent slopes, $k_c$ and $k_o$, because the quality of ATLAS data is not sufficient to see apparent phase effects for different colors. However, with more accurate data, it might be necessary to include this effect in the modeling, which might be the case of future LSST observations in different filters \citep{Hsi.ea:19}. 

    \subsection{Colors}

\begin{figure*}[t]
        \includegraphics[width=\columnwidth]{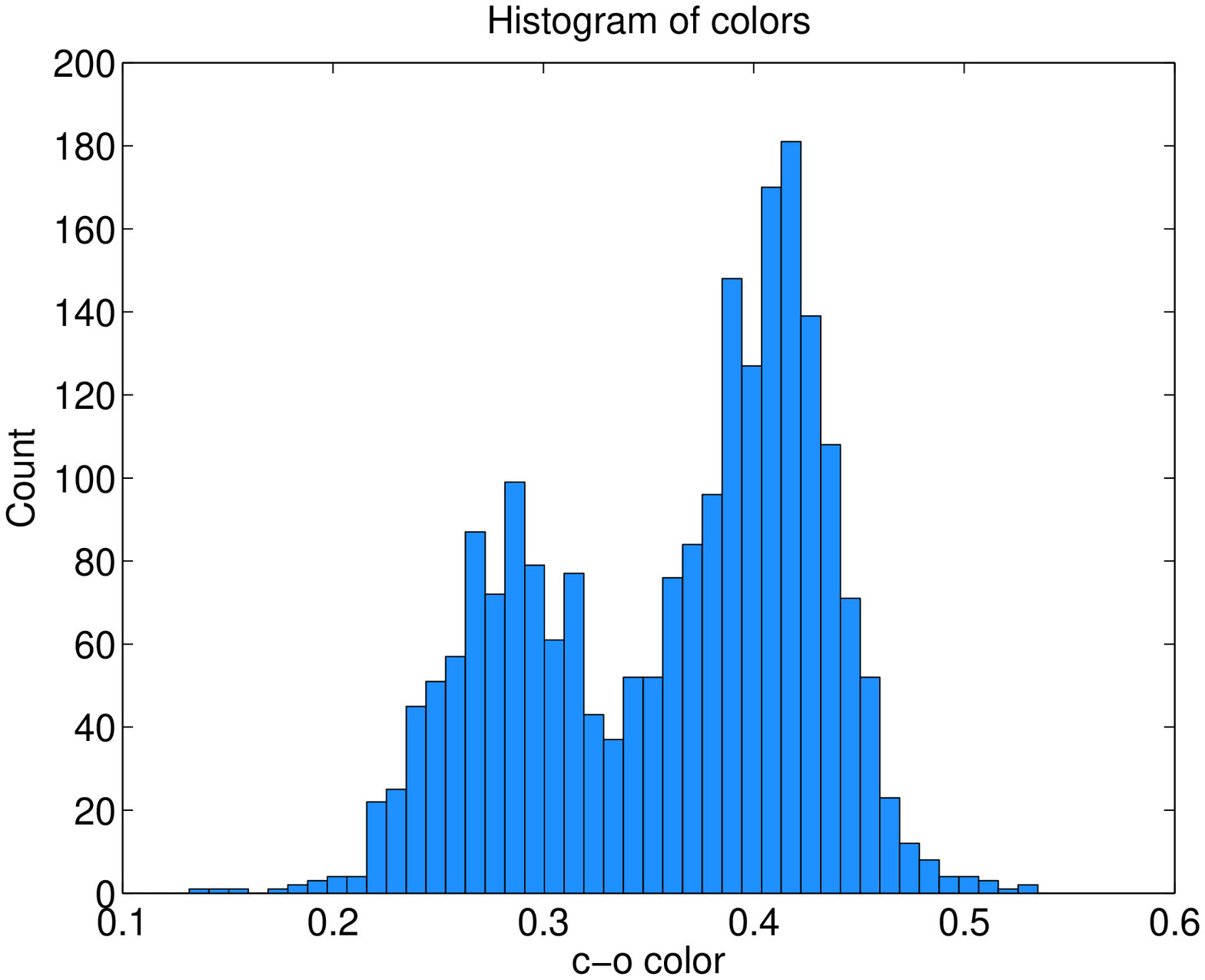}
        \includegraphics[width=\columnwidth]{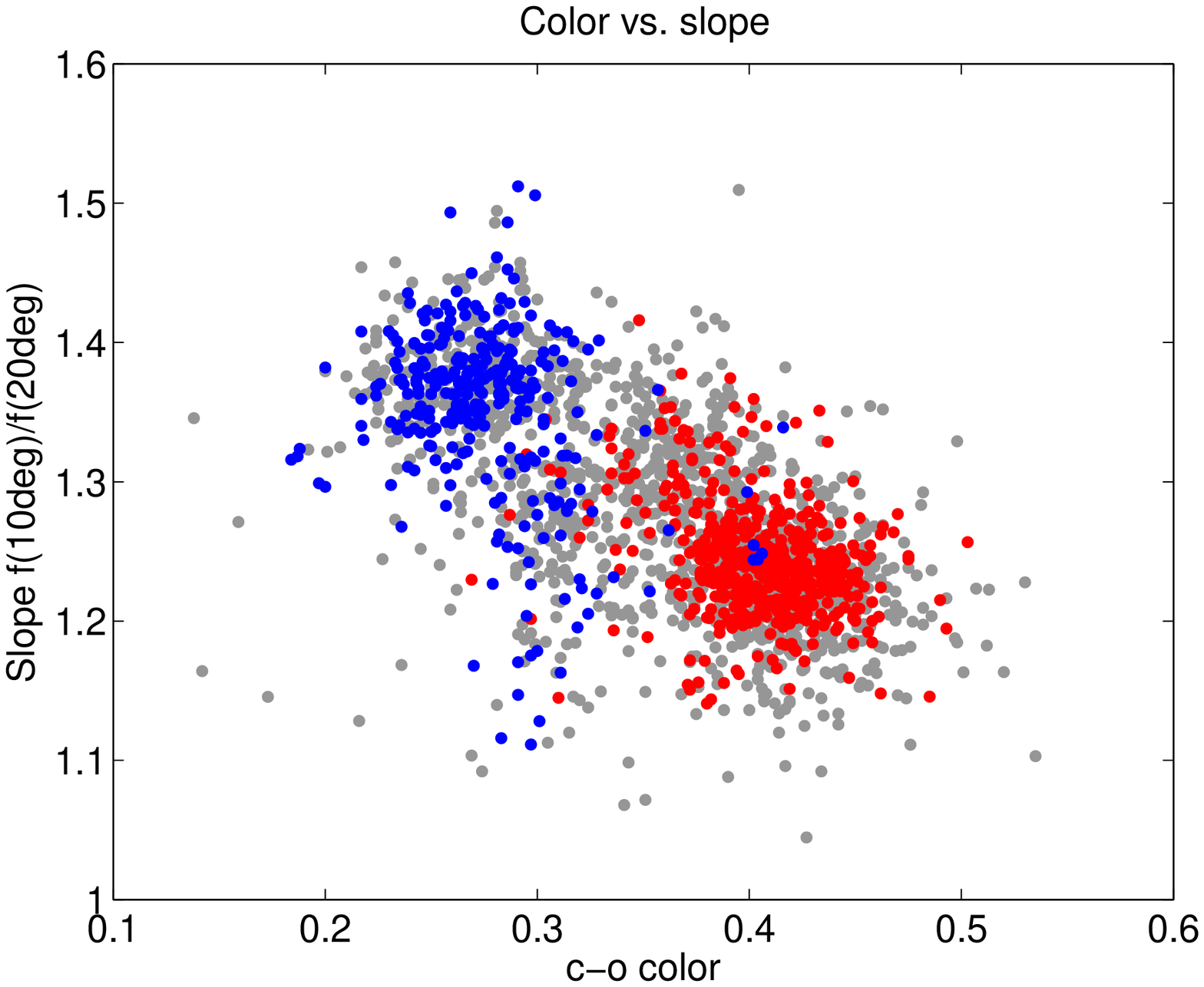}

        \includegraphics[width=\columnwidth]{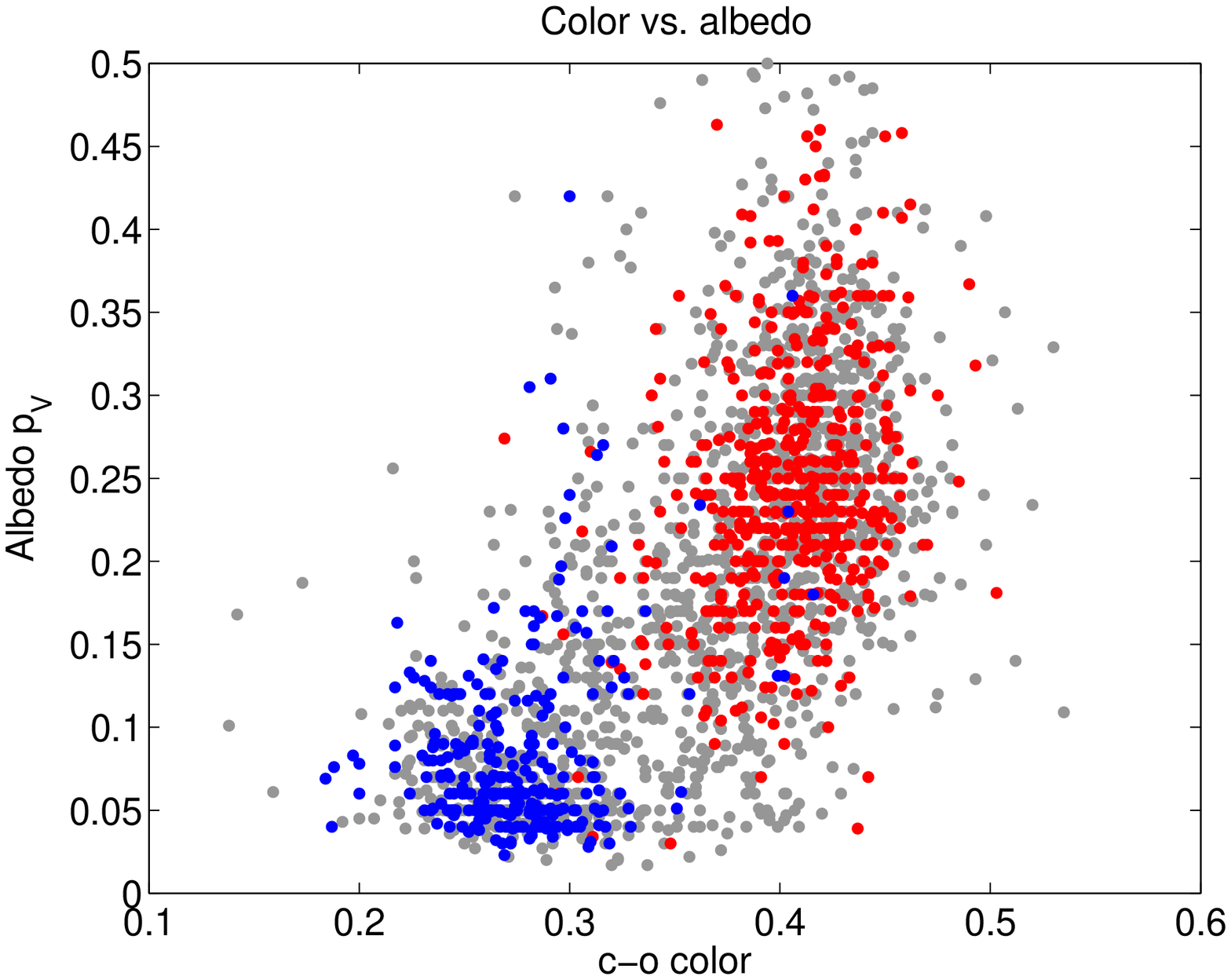}
        \includegraphics[width=\columnwidth]{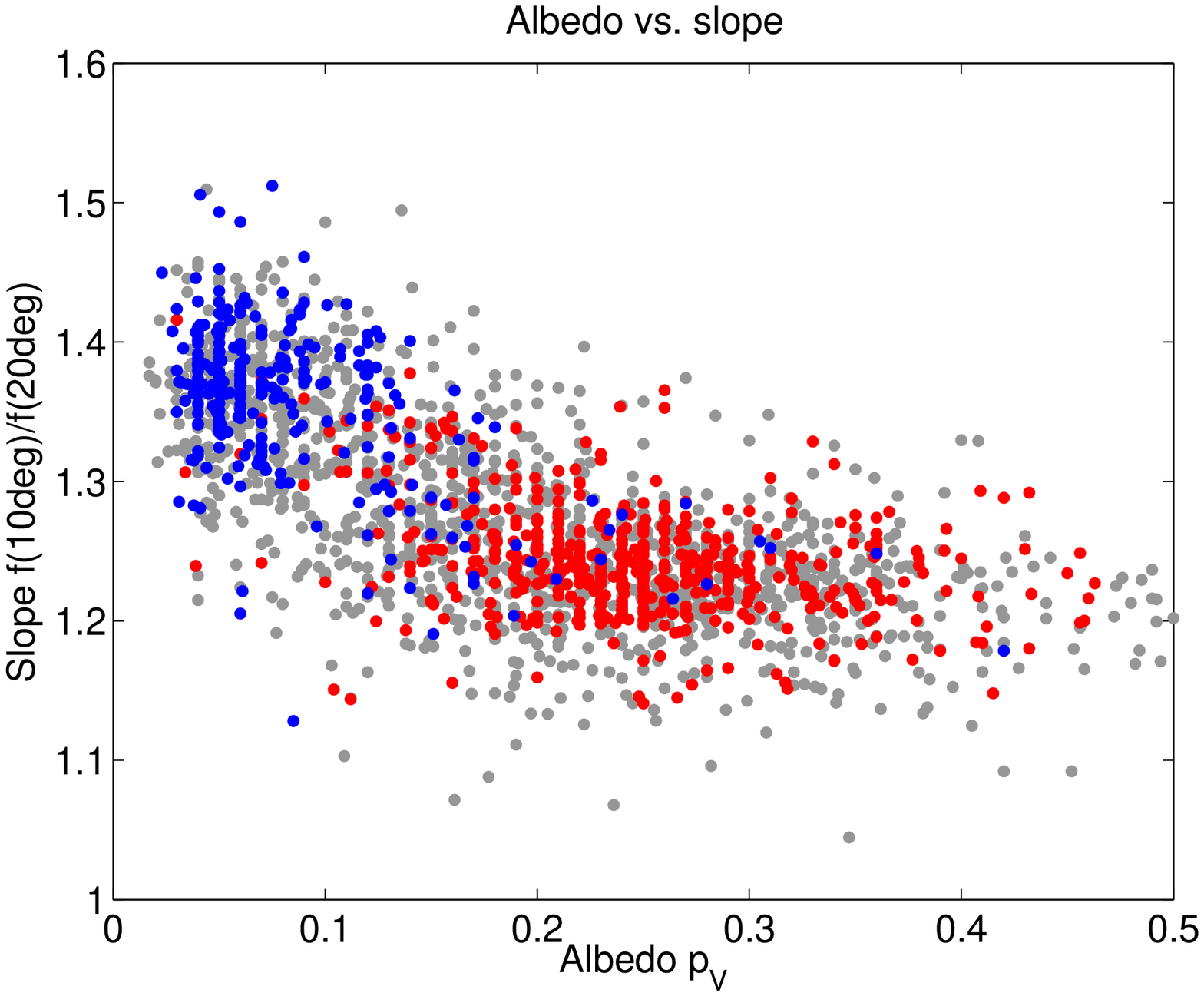}
        \caption{Distribution of colors $c - o$ and their correlations with albedo $p_V$ and slope of the phase function. Red points are S-complex asteroids, blue points are C-complex asteroids, and gray points are the remaining asteroids with a different or unknown taxonomy. Information about taxonomy and albedo was taken from the database of \cite{Del.ea:17}.}
        \label{fig:correlations}
      \end{figure*}

      There were observations in two filters for almost all asteroids in our sample so we could determine the color index $c - o$ accurately. The color index for each asteroid is listed in Tables~\ref{tab:models} to \ref{tab:models_interval_partial}. The histogram of colors for all of the models we derived, shown in Fig.~\ref{fig:correlations}, is clearly bimodal, and the color is correlated with taxonomic class. The same figure also shows correlations between other parameters as follows: the slope of the phase curve, albedo, and taxonomic class. We defined the slope of the phase function as a ratio of values at phase angles $10^\circ$ and $20^\circ$ according to eq.~(\ref{eq:phase_function}): $f(10^\circ) / f(20^\circ)$. Taxonomy information and albedo values were taken from the compilation database of \cite{Del.ea:17}.\footnote{\url{https://www-n.oca.eu/delbo/astphys/astphys.html}} To show correlations, we divided asteroids into C-complex (B, C, Cb, Cg, Cgh, and Ch) and S-complex (S, Sa, Sq, Sr, and Sv) according to Bus-DeMeo taxonomy \citep{DeM.ea:09}. We can see the expected trends: S-complex asteroids with a higher albedo have a higher color index and smaller slopes than low-albedo C-complex asteroids. 

    \subsection{Slow rotators}

      Recent analyses of asteroid photometry from Kepler \citep{Kis.ea:19} and TESS \citep{McN.ea:19, Pal.ea:20} missions show that the population of slow rotators is significantly larger than previously estimated from the statistics of periods obtained from ground-based observations. The bias against slow rotators \citep{Mar.ea:15, Mar.ea:18} is a consequence of practical problems of composing a full, long-period light curve from individual short parts observed in separate nights. However, accurate sparse photometry covering several years of observations should not be strongly biased against slow rotators. Indeed, we were able to successfully reconstruct shape models for many asteroids with the rotation period longer than one day.  
      
\begin{figure}[t]
        \includegraphics[width=\columnwidth]{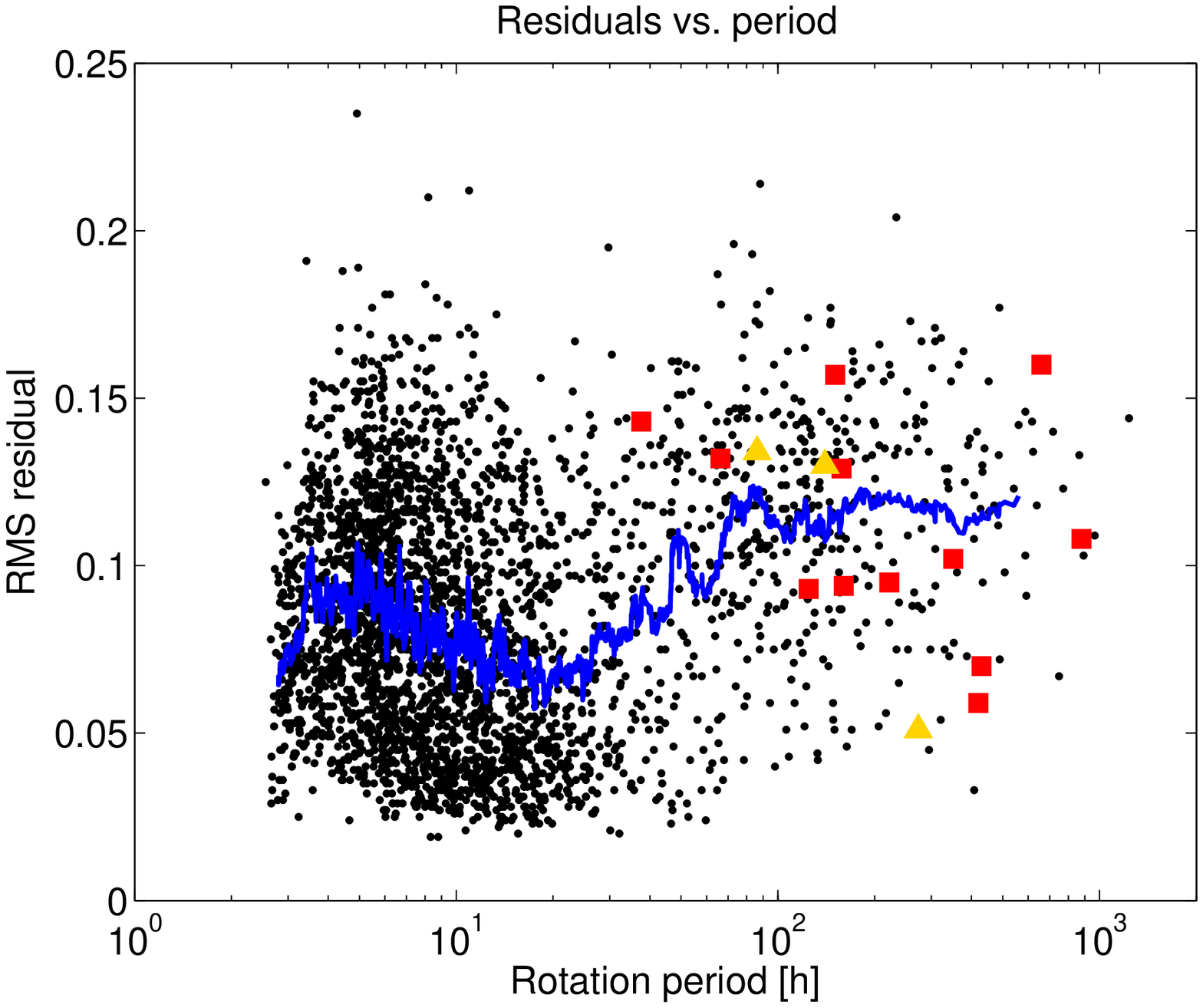}
        \caption{Dependence of rms residual on rotation period. Black points represent individual models, and the blue curve is a running mean of residuals over 31 points centered at the middle period. Red squares mark known tumblers (T or T+ tag in the Lightcurve Database of \cite{War.ea:09}), and yellow triangles mark asteroids that are known to be in the principal-axis rotation state, i.e., not tumbling \mbox{(T--)}.}
        \label{fig:rms_vs_period}
      \end{figure}

      The dependence of the rms residual on the rotation period for all the models that we derived is shown in Fig.~\ref{fig:rms_vs_period}. The values range from 0.02\,mag for the best fits to about 0.2\,mag for the worst ones. On average, slow rotators have larger residuals than the rest of the population. For example, the mean residual for asteroids with $P < 20$\,h is 0.08\,mag; whereas, for slow rotators with $P > 100$\,h, it is 0.12\,mag (about the same values also apply for the medians). This difference is statistically significant since the t-test on these two samples reveals a practically zero probability to which our derived means are equal. The difference may be caused by some unknown dependencies of the inversion process on the rotation period. However, another speculative explanation might be that slow rotators are actually not rotating in an exact principal-axis mode, as our model assumes, but they are slightly excited. This would mean that we can fit the data with a single-period relaxed model, but the residuals are somewhat higher because the real light curve is more complicated than the model can describe. However, the residuals for known tumblers and slow non-tumbling asteroids, according to the LCDB shown in Fig.~\ref{fig:rms_vs_period}, do not follow this trend. Although, the number of these asteroids in our sample is still low. We do not have any explanation for the minimum in rms at $P \approx 20$\,h, but this might be caused by some 24\,h bias.

  \section{Conclusions}

    The ATLAS photometric data of asteroids can be successfully used for their shape and spin reconstruction. Although the success rate of unique model determination from the current data set is not very high (less than 3000 models from about 100,000 asteroids), the ATLAS survey continues to observe every night. And we estimate that by including another year of observations (from October 2018 to the present day), we could derive hundreds or thousands of other models. By continuously adding and processing ATLAS data, we can create new models and update existing ones. On top of that, we can combine ATLAS with photometry from other surveys, which would be more efficient in producing asteroid models than inverting different data sets separately. The most promising possibility in the near future is to jointly invert photometry from Gaia Data Release 3, which is scheduled for the second half of 2021, and ATLAS.
    
  \begin{acknowledgements}
    JD and RV were supported by the grant 18-04514J of the Czech Science Foundation. We greatly appreciate the contribution of tens of thousands of volunteers who joined the Asteroids@home BOINC project and provided their computing resources.  This work has made use of data from the Asteroid Terrestrial-impact Last Alert System (ATLAS) project. ATLAS is primarily funded to search for near earth asteroids through NASA grants NN12AR55G, 80NSSC18K0284, and 80NSSC18K1575; byproducts of the NEO search include images and
  catalogs from the survey area.  The ATLAS science products have been made possible through the contributions of the University of Hawaii Institute for Astronomy, the Queen's University Belfast, the Space Telescope Science Institute, and the South African Astronomical Observatory. This research has made use of IMCCE's Miriade VO tool. This research was supported by the Munich Institute for Astro- and Particle Physics (MIAPP) of the DFG cluster of excellence ``Origin and Structure of the Universe.''
  \end{acknowledgements}

  \input{ATLAS.bbl}

  \clearpage

\begin{appendix}
   
  \section{List of new models}

  \onecolumn
  
  \begin{landscape}
  \longtab{
    \tiny
\begin{longtable}{r l r c r c r c r c @{} d l @{} d l c r r c c c}
      \caption{\label{tab:models} List of new asteroid models. For each asteroid, we list one or two pole directions in the ecliptic coordinates $(\lambda, \beta)$ with uncertainties $\sigma_\lambda$ and $\sigma_\beta$, the sidereal rotation period $P$ and its error $\sigma_P$, the rotation period $P_\mathrm{LCDB}$ from the LCDB and its quality code $U$ (if available), the number of sparse photometric data points in orange  and cyan filters $N_o$,  $N_c$, respectively, the color $c - o$ and its error $\sigma_{c-o}$, and the method which was used to derive the rotation period: C -- convex models, E -- ellipsoids, CE -- both methods gave the same unique period. The accuracy of the sidereal rotation period $P$ is of the order of the last decimal place given. For some asteroids, the LCDB contains a note about the rotation state or binarity: Here A means an ambiguous period; B stands for a binary system; T means tumbling -- a non-principal-axis rotation; T0 means that the theoretical tumbling damping time scale \citep{Pra.ea:14} is long enough that tumbling might be expected, but observations are not sufficient to substantiate either tumbling or not tumbling; T$-$ means that the tumbling damping time scale is long enough that tumbling might be expected, but observations indicate that the object is not tumbling; and T+ indicates that the tumbling damping time scale is short enough that tumbling would not seem likely, however observations indicate that it may be tumbling or actually is tumbling.}\\
      \hline\hline
      \multicolumn{2}{c}{Asteroid}		& \multicolumn{1}{c}{$\lambda_1$}	& \multicolumn{1}{c}{$\sigma_{\lambda_1}$}	& \multicolumn{1}{c}{$\beta_1$}	& \multicolumn{1}{c}{$\sigma_{\beta_1}$}	& \multicolumn{1}{c}{$\lambda_2$}	& \multicolumn{1}{c}{$\sigma_{\lambda_2}$}	& \multicolumn{1}{c}{$\beta_2$}	& \multicolumn{1}{c}{$\sigma_{\beta_2}$}	& \multicolumn{1}{c}{$P$}	& \multicolumn{1}{c}{$\sigma_P$}	& \multicolumn{1}{c}{$P_\mathrm{LCDB}$}	& \multicolumn{1}{c}{U}	&	Note	& \multicolumn{1}{c}{$N_o$}	& \multicolumn{1}{c}{$N_c$}	& $c - o$	& $\sigma_{c - o}$	& method	\\
      number	& name/designation		& \multicolumn{1}{c}{[deg]}		& \multicolumn{1}{c}{[deg]}			& \multicolumn{1}{c}{[deg]}	& \multicolumn{1}{c}{[deg]}			& \multicolumn{1}{c}{[deg]}		& \multicolumn{1}{c}{[deg]}			& \multicolumn{1}{c}{[deg]}	& \multicolumn{1}{c}{[deg]}			& \multicolumn{1}{c}{[h]}	& \multicolumn{1}{c}{[h]} 		& \multicolumn{1}{c}{[h]}		& 			& 		&				&				& [mag]		&			&		\\
      \hline
      \endfirsthead
      \caption{continued.}\\
      \hline\hline
      \multicolumn{2}{c}{Asteroid}		& \multicolumn{1}{c}{$\lambda_1$}	& \multicolumn{1}{c}{$\sigma_{\lambda_1}$}	& \multicolumn{1}{c}{$\beta_1$}	& \multicolumn{1}{c}{$\sigma_{\beta_1}$}	& \multicolumn{1}{c}{$\lambda_2$}	& \multicolumn{1}{c}{$\sigma_{\lambda_2}$}	& \multicolumn{1}{c}{$\beta_2$}	& \multicolumn{1}{c}{$\sigma_{\beta_2}$}	& \multicolumn{1}{c}{$P$}	& \multicolumn{1}{c}{$\sigma_P$}	& \multicolumn{1}{c}{$P_\mathrm{LCDB}$}	& \multicolumn{1}{c}{U}	&	Note	& \multicolumn{1}{c}{$N_o$}	& \multicolumn{1}{c}{$N_c$}	& $c - o$	& $\sigma_{c - o}$	& method	\\
      number	& name/designation		& \multicolumn{1}{c}{[deg]}		& \multicolumn{1}{c}{[deg]}			& \multicolumn{1}{c}{[deg]}	& \multicolumn{1}{c}{[deg]}			& \multicolumn{1}{c}{[deg]}		& \multicolumn{1}{c}{[deg]}			& \multicolumn{1}{c}{[deg]}	& \multicolumn{1}{c}{[deg]}			& \multicolumn{1}{c}{[h]}	& \multicolumn{1}{c}{[h]} 		& \multicolumn{1}{c}{[h]}		& 			& 		&				&				& [mag]		&			&		\\
      \hline
      \endhead
      \hline
      \endfoot
      \input{spin_results_ATLAS_errors}
\end{longtable}
  }
  
  \newpage

  \longtab{
    \tiny
    \begin{longtable}{r l r c r c r c r c @{} d l @{} d l c r r c c c}
      \caption{\label{tab:models_interval} List of new asteroid models with period searched on a limited interval. The meaning of columns is the same as in Table~\ref{tab:models}}\\
      \hline\hline
      \multicolumn{2}{c}{Asteroid}		& \multicolumn{1}{c}{$\lambda_1$}	& \multicolumn{1}{c}{$\sigma_{\lambda_1}$}	& \multicolumn{1}{c}{$\beta_1$}	& \multicolumn{1}{c}{$\sigma_{\beta_1}$}	& \multicolumn{1}{c}{$\lambda_2$}	& \multicolumn{1}{c}{$\sigma_{\lambda_2}$}	& \multicolumn{1}{c}{$\beta_2$}	& \multicolumn{1}{c}{$\sigma_{\beta_2}$}	& \multicolumn{1}{c}{$P$}	& \multicolumn{1}{c}{$\sigma_P$}	& \multicolumn{1}{c}{$P_\mathrm{LCDB}$}	& \multicolumn{1}{c}{U}	&	Note	& \multicolumn{1}{c}{$N_o$}	& \multicolumn{1}{c}{$N_c$}	& $c - o$	& $\sigma_{c - o}$	& method	\\
      number	& name/designation		& \multicolumn{1}{c}{[deg]}		& \multicolumn{1}{c}{[deg]}			& \multicolumn{1}{c}{[deg]}	& \multicolumn{1}{c}{[deg]}			& \multicolumn{1}{c}{[deg]}		& \multicolumn{1}{c}{[deg]}			& \multicolumn{1}{c}{[deg]}	& \multicolumn{1}{c}{[deg]}			& \multicolumn{1}{c}{[h]}	& \multicolumn{1}{c}{[h]} 		& \multicolumn{1}{c}{[h]}		& 			& 		&				&				& [mag]		&			&		\\
      \hline
      \endfirsthead
      \caption{continued.}\\
      \hline\hline
      \multicolumn{2}{c}{Asteroid}		& \multicolumn{1}{c}{$\lambda_1$}	& \multicolumn{1}{c}{$\sigma_{\lambda_1}$}	& \multicolumn{1}{c}{$\beta_1$}	& \multicolumn{1}{c}{$\sigma_{\beta_1}$}	& \multicolumn{1}{c}{$\lambda_2$}	& \multicolumn{1}{c}{$\sigma_{\lambda_2}$}	& \multicolumn{1}{c}{$\beta_2$}	& \multicolumn{1}{c}{$\sigma_{\beta_2}$}	& \multicolumn{1}{c}{$P$}	& \multicolumn{1}{c}{$\sigma_P$}	& \multicolumn{1}{c}{$P_\mathrm{LCDB}$}	& \multicolumn{1}{c}{U}	&	Note	& \multicolumn{1}{c}{$N_o$}	& \multicolumn{1}{c}{$N_c$}	& $c - o$	& $\sigma_{c - o}$	& method	\\
      number	& name/designation		& \multicolumn{1}{c}{[deg]}		& \multicolumn{1}{c}{[deg]}			& \multicolumn{1}{c}{[deg]}	& \multicolumn{1}{c}{[deg]}			& \multicolumn{1}{c}{[deg]}		& \multicolumn{1}{c}{[deg]}			& \multicolumn{1}{c}{[deg]}	& \multicolumn{1}{c}{[deg]}			& \multicolumn{1}{c}{[h]}	& \multicolumn{1}{c}{[h]} 		& \multicolumn{1}{c}{[h]}		& 			& 		&				&				& [mag]		&			&		\\
      \hline
      \endhead
      \hline
      \endfoot
      \input{spin_results_ATLAS_interval_errors}
\end{longtable}
  }
  \end{landscape}

  \newpage

  \longtab{
    \tiny
    \begin{longtable}{r l r r @{} d l @{} d l c r r c c c}
      \caption{\label{tab:models_partial} List of partial models with the mean ecliptic pole latitude $\beta$ and its dispersion $\Delta\beta$. The meaning of other columns is the same as in Table~\ref{tab:models}}\\
      \hline\hline
      \multicolumn{2}{c}{Asteroid}		& \multicolumn{1}{c}{$\beta$}	& \multicolumn{1}{c}{$\Delta\beta$}	& \multicolumn{1}{c}{$P$}	& \multicolumn{1}{c}{$\sigma_P$}	& \multicolumn{1}{c}{$P_\mathrm{LCDB}$}	& \multicolumn{1}{c}{U}	&	Note	& \multicolumn{1}{c}{$N_o$}	& \multicolumn{1}{c}{$N_c$}	& $c - o$	& $\sigma_{c - o}$	& method	\\
      number	& name/designation		& \multicolumn{1}{c}{[deg]}	& \multicolumn{1}{c}{[deg]}	& \multicolumn{1}{c}{[h]}	& \multicolumn{1}{c}{[h]}		& \multicolumn{1}{c}{[h]} 		& 			& 		&				&				& [mag]		&			&		\\
      \hline
      \endfirsthead
      \caption{continued.}\\
      \hline\hline
      \multicolumn{2}{c}{Asteroid}		& \multicolumn{1}{c}{$\beta$}	& \multicolumn{1}{c}{$\Delta\beta$}	& \multicolumn{1}{c}{$P$}	& \multicolumn{1}{c}{$\sigma_P$}	& \multicolumn{1}{c}{$P_\mathrm{LCDB}$}	& \multicolumn{1}{c}{U}	&	Note	& \multicolumn{1}{c}{$N_o$}	& \multicolumn{1}{c}{$N_c$}	& $c - o$	& $\sigma_{c - o}$	& method	\\
      number	& name/designation		& \multicolumn{1}{c}{[deg]}	& \multicolumn{1}{c}{[deg]}	& \multicolumn{1}{c}{[h]}	& \multicolumn{1}{c}{[h]}		& \multicolumn{1}{c}{[h]} 		& 			& 		&				&				& [mag]		&			&		\\
      \hline
      \endhead
      \hline
      \endfoot
      \input{spin_results_ATLAS_beta_errors}
\end{longtable}
  }
  
  \newpage

  \longtab{
    \tiny
    \begin{longtable}{r l r r @{} d l @{} d l c r r c c c}
      \caption{\label{tab:models_interval_partial} List of partial models with the mean ecliptic pole latitude $\beta$ and its dispersion $\Delta\beta$. The meaning of other columns is the same as in Table~\ref{tab:models}}\\
      \hline\hline
      \multicolumn{2}{c}{Asteroid}		& \multicolumn{1}{c}{$\beta$}	& \multicolumn{1}{c}{$\Delta\beta$}	& \multicolumn{1}{c}{$P$}	& \multicolumn{1}{c}{$\sigma_P$}	& \multicolumn{1}{c}{$P_\mathrm{LCDB}$}	& \multicolumn{1}{c}{U}	&	Note	& \multicolumn{1}{c}{$N_o$}	& \multicolumn{1}{c}{$N_c$}	& $c - o$	& $\sigma_{c - o}$	& method	\\
      number	& name/designation		& \multicolumn{1}{c}{[deg]}	& \multicolumn{1}{c}{[deg]}	& \multicolumn{1}{c}{[h]}	& \multicolumn{1}{c}{[h]}		& \multicolumn{1}{c}{[h]} 		& 			& 		&				&				& [mag]		&			&		\\
      \hline
      \endfirsthead
      \caption{continued.}\\
      \hline\hline
      \multicolumn{2}{c}{Asteroid}		& \multicolumn{1}{c}{$\beta$}	& \multicolumn{1}{c}{$\Delta\beta$}	& \multicolumn{1}{c}{$P$}	& \multicolumn{1}{c}{$\sigma_P$}	& \multicolumn{1}{c}{$P_\mathrm{LCDB}$}	& \multicolumn{1}{c}{U}	&	Note	& \multicolumn{1}{c}{$N_o$}	& \multicolumn{1}{c}{$N_c$}	& $c - o$	& $\sigma_{c - o}$	& method	\\
      number	& name/designation		& \multicolumn{1}{c}{[deg]}	& \multicolumn{1}{c}{[deg]}	& \multicolumn{1}{c}{[h]}	& \multicolumn{1}{c}{[h]}		& \multicolumn{1}{c}{[h]} 		& 			& 		&				&				& [mag]		&			&		\\
      \hline
      \endhead
      \hline
      \endfoot
      \input{spin_results_ATLAS_interval_beta_errors}
\end{longtable}
  }
  
  \end{appendix}

\end{document}

%% file: ATLAS.bbl
\newcommand{\SortNoop}[1]{}

%% file: spin_results_ATLAS_errors.tex
     33 & Polyhymnia                &     9 &   2 & $-52$ &   3 &   187 &   2 & $-53$ &   3 &  18.6092 &   0.0001 &      18.608 &     3 &         &   659 &  173 &  0.368 & 0.004 &   C  \\
     84 & Klio                      &   356 &   5 & $-53$ &   6 &       &     &       &     &  23.5752 &   0.0007 &      23.562 &     3 &         &   299 &  161 &  0.284 & 0.005 &  CE  \\
     91 & Aegina                    &    40 &   4 &    32 &   3 &   218 &   5 &    36 &   4 &  6.02819 &  0.00003 &       6.025 &     3 &         &   310 &   94 &  0.278 & 0.006 &   C  \\
    123 & Brunhild                  &    78 &   3 &    51 &   4 &   242 &   2 &    38 &   5 &  9.87346 &  0.00004 &       9.873 &     3 &         &   401 &  174 &  0.378 & 0.004 &   C  \\
    148 & Gallia                    &   140 &   2 & $-17$ &   3 &       &     &       &     &  20.6636 &   0.0004 &      20.664 &     3 &         &   395 &   85 &  0.380 & 0.008 &   C  \\
    151 & Abundantia                &     3 &   2 &    16 &   3 &   188 &   1 &    32 &   6 &  9.86432 &  0.00008 &       9.864 &     3 &         &   297 &  113 &  0.442 & 0.006 &   C  \\
    156 & Xanthippe                 &    22 &   2 &    15 &   4 &   197 &   3 &     9 &   6 &  22.1157 &   0.0004 &       22.37 &     3 &         &   395 &  118 &  0.270 & 0.005 &   E  \\
    172 & Baucis                    &     6 &   3 & $-34$ &   5 &   203 &   4 & $-52$ &   4 &   27.410 &    0.001 &      27.417 &     3 &         &   315 &  160 &  0.434 & 0.005 &   E  \\

%% file: spin_results_ATLAS_interval_errors.tex
     67 & Asia                      &   117 &   2 &    42 &   4 &   286 &   3 &    50 &   4 &  15.8504 &   0.0003 &      15.853 &     3 &         &   224 &  101 &  0.404 & 0.007 &   E  \\
     86 & Semele                    &    44 &   4 & $-43$ &   4 &   213 &   3 & $-31$ &   3 &  16.6420 &   0.0002 &      16.634 &     3 &         &   445 &  142 &  0.264 & 0.004 &   C  \\
    118 & Peitho                    &   176 &   1 &    59 &   4 &   346 &   1 &    34 &   3 &  7.80648 &  0.00004 &      7.8055 &     3 &         &   385 &  174 &  0.402 & 0.004 &  CE  \\
    124 & Alkeste                   &   136 &   3 &    16 &   6 &   313 &   3 &    11 &   4 &  9.90702 &  0.00007 &       9.906 &     3 &         &   203 &  122 &  0.404 & 0.005 &   E  \\
    204 & Kallisto                  &    15 &   5 &    30 &   5 &   188 &   5 &    14 &   8 &  19.4870 &   0.0004 &      19.489 &     3 &         &   368 &  127 &  0.390 & 0.005 &   E  \\
    236 & Honoria                   &    14 &   3 & $ -5$ &   7 &   195 &   3 &     9 &   8 &  12.3399 &   0.0003 &      12.336 &  3$-$ &         &   211 &   92 &  0.399 & 0.008 &   C  \\
    331 & Etheridgea                &    75 &   2 &    60 &   4 &   264 &   2 &    47 &   6 &  25.3141 &   0.0003 &      25.315 &     3 &         &   453 &  156 &  0.278 & 0.004 &  CE  \\
    374 & Burgundia                 &    13 &   1 &    41 &   4 &   180 &   1 &    23 &   3 &  6.96394 &  0.00005 &       6.972 &     3 &         &   398 &  124 &  0.398 & 0.005 &  CE  \\

%% file: spin_results_ATLAS_beta_errors.tex
     61 & Danae                     & $-38$ &  20 &  11.5499 &   0.0003 &       11.45 &     3 &         &   366 &  165 &  0.399 & 0.005 &  CE  \\
    113 & Amalthea                  & $-19$ &  25 &  9.93936 &  0.00004 &       9.950 &     3 &         &   292 &  146 &  0.405 & 0.006 &   C  \\
    133 & Cyrene                    & $-48$ &  18 &  12.7087 &   0.0001 &      12.708 &     3 &         &   311 &  166 &  0.428 & 0.005 &  CE  \\
    139 & Juewa                     & $ -4$ &  12 &  20.9849 &   0.0003 &      20.991 &     3 &         &   309 &   86 &  0.283 & 0.006 &   C  \\
    325 & Heidelberga               &    39 &  21 &  6.73597 &  0.00003 &       6.737 &     3 &         &   374 &  127 &  0.308 & 0.005 &   C  \\
    358 & Apollonia                 & $-18$ &  24 &  25.3108 &   0.0005 &        50.6 &  3$-$ &         &   359 &  164 &  0.276 & 0.005 &   C  \\
    370 & Modestia                  & $-50$ &   9 &  22.5411 &   0.0003 &     22.5299 &     3 &         &   339 &  133 &  0.304 & 0.006 &  CE  \\
    420 & Bertholda                 &    64 &  14 & 10.98717 &  0.00004 &       11.04 &     3 &         &   319 &  164 &  0.328 & 0.003 &  CE  \\

%% file: spin_results_ATLAS_interval_beta_errors.tex
     59 & Elpis                     & $-43$ &  14 &  13.6741 &   0.0001 &      13.671 &     3 &         &   172 &  114 &  0.265 & 0.005 &   E  \\
     74 & Galatea                   &     5 &  11 &  17.2678 &   0.0001 &      17.268 &     3 &         &   423 &  153 &  0.270 & 0.004 &   C  \\
    168 & Sibylla                   & $-44$ &  14 &   47.014 &    0.002 &      47.009 &     3 &         &   490 &  153 &  0.279 & 0.003 &  CE  \\
    228 & Agathe                    & $-60$ &  14 &  6.48653 &  0.00004 &       6.484 &     3 &         &   322 &  118 &  0.450 & 0.007 &   C  \\
    229 & Adelinda                  & $ -1$ &  22 &  6.59934 &  0.00004 &        6.60 &     3 &         &   398 &  166 &  0.249 & 0.007 &   C  \\
    303 & Josephina                 & $-32$ &   7 &  12.4996 &   0.0001 &      12.497 &     3 &         &   400 &  123 &  0.296 & 0.005 &   C  \\
    375 & Ursula                    &     9 &  18 &  16.8937 &   0.0003 &      16.899 &     3 &         &   341 &  168 &  0.262 & 0.004 &   E  \\
    422 & Berolina                  & $-60$ &  11 &  25.9764 &   0.0003 &      25.978 &     3 &         &   303 &  129 &  0.290 & 0.006 &   E  \\